\documentclass[preprint,12pt, a4paper]{elsarticle}

\usepackage{amssymb}
\usepackage{hyperref}
\usepackage{float}
\usepackage{natbib}
\usepackage{listings}
\usepackage{color}
\usepackage{xcolor}
\usepackage{multicol}
\usepackage{enumitem}

\usepackage{caption}
\DeclareCaptionFont{white}{\color{white}}
\DeclareCaptionFormat{listing}{\colorbox{gray}{\parbox{\textwidth}{#1#2#3}}}
\journal{Software Impacts}

\begin{document}

\begin{frontmatter}



\title{DRAS-CQSim: A Reinforcement Learning based Framework for HPC Cluster Scheduling}


\author{Yuping Fan, Zhiling Lan}

\address{Illinois Institute of Technology, Chicago, IL}

\begin{abstract}
For decades, system administrators have been striving to design and tune cluster scheduling policies to improve the performance of high performance computing (HPC) systems. However, the increasingly complex HPC systems combined with highly diverse workloads make such manual process challenging, time-consuming, and error-prone. We present a reinforcement learning based HPC scheduling framework named DRAS-CQSim to automatically learn optimal scheduling policy. DRAS-CQSim encapsulates simulation environments, agents, hyperparameter tuning options, and different reinforcement learning algorithms, which allows the system administrators to quickly obtain customized scheduling policies.
\end{abstract}


\end{frontmatter}



\section{Introduction}
Cluster scheduler is crucial in HPC. It determines when and which user jobs should be allocated to available system resources. Traditionally, scheduling policies are developed by system administrators based on their experience with specific systems and workloads. However, such a manually designing and tuning process becomes increasingly difficult to handle the increasingly complex HPC systems and highly diverse application workloads. 
In recent years, reinforcement learning has been successfully employed in various fields of decision-making problems, such as gaming playing \cite{Go2016,Go2017}, self-driving cars \cite{shalevshwartz2016safe,Ahmad2017}, and autonomous robots \cite{Smart2002,Kober2014,Johannink2019}. Reinforcement learning (RL) is an area of machine learning that automatically learns to make decisions through interaction with the environment. 

Inspired by the successful RL examples, we present an open-source HPC scheduling toolkit named DRAS (Deep Reinforcement Agent for Scheduling) to automatically learn customized scheduling policies \cite{Fan5, DRASGithub}. DRAS with CQSim simulator packs together all the necessary components, i.e., training environment, agents, and RL algorithms, to train the scheduling policy model. Currently DRAS contains the two most popular reinforcement learning algorithms, deep q-learning, and policy gradient \cite{Sutton1999}, and can be easily switch to other reinforcement learning or traditional scheduling algorithms. Our objective is to enable system administrators to quickly obtain the optimal scheduling policies for their specific system and workload environment and easily compare the performance of different scheduling policies.
 

\begin{figure}[htbp]
\centerline{\includegraphics[width=0.9\linewidth]{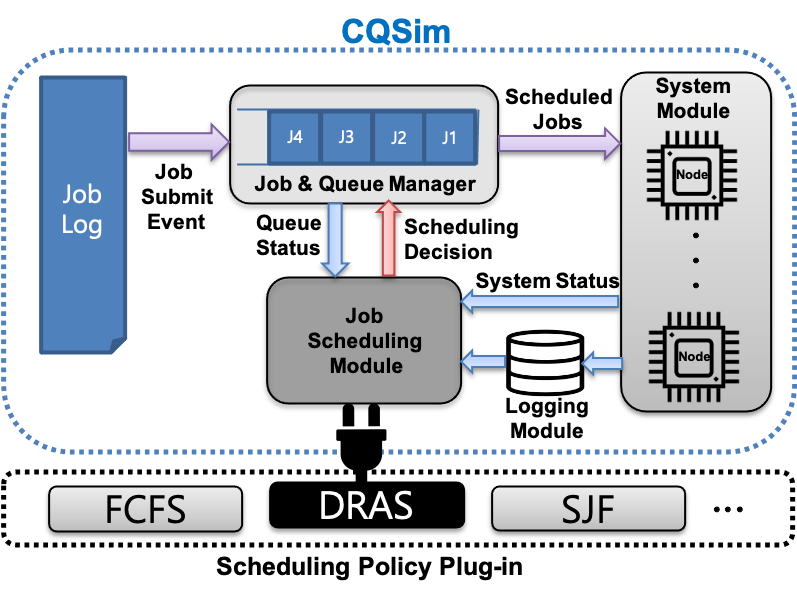}}
\caption{DRAS-CQSim overview.}
\label{fig1}
\end{figure}

\section{Functionalities and Key Features}\label{Functionalities and Key Features}
DRAS-CQSim, illustrated in Figure \ref{fig1}, is a reinforcement learning empowered cluster scheduling framework. Rather than executing jobs on real systems, DRAS uses the event-driven scheduling simulator named CQSim \cite{CQSimGithub} to train the agents. CQSim simulates the job scheduling environment by reading the job arrival event from the job log and advancing the simulation clock according to scheduling decisions and job runtime information. CQSim consists of four main components: job \& queue manager, system module, logging module, and job scheduling module. The job \& queue manager maintains waiting jobs and manages job lifecycle. System module simulates the status of the real systems. Each node in the system is represented as an object and the system module keeps track of each node's availability information. The logging module collects the finished job information, such as job submit time, start time, and end time. The job scheduling module makes scheduling decisions based on the queue status, system status, and historical job information retrieved from queue manager, system module, and logging module respectively. The job scheduling module provides an interface to plugin a customized scheduling policy, such as FCFS (First Come First Serve) and SJF (Shortest Job First). Our DRAS agents are implemented as reinforcement learning based scheduling policies. Our key features are:

\begin{itemize}[leftmargin=*]
\item Easy-to-use: DRAS-CQSim requires Python, Tensorflow, and Keras to be install. These prerequisites are easily satisfied on most systems. It provides the concise command to run simulations and train agents.  Listing \ref{basic_command} shows an example to train DRAS agents. Argument \verb|n| and \verb|j| are required to run DRAS. Argument \verb|n| specifies the file containing the information of the simulated system. Argument \verb|j| specifies the file containing job traces. The \verb|Config| folder encapsulates all details of scheduling policies, simulated systems, neural networks, requiring no configuration from the user for most use cases. DRAS provides sensible default values, but users can customize configurations via two approaches: directly modify the configuration files under \verb|Config| folder or add optional arguments in command. 
\begin{lstlisting}[language=command.com, caption=Train DRAS agent command example,label=basic_command]
$ python cqsim.py -j job_log.swf -n node_structure.swf \
         --is_training 1
\end{lstlisting}
\item High scalability: The simulator dynamically streams in the job traces for simulation and streams out the finished jobs to disks. Hence, the memory requirements to run the simulations do not linearly increase with the size of job traces.  This allows the simulator to run scalable simulations with a limited amount of memory resources. 
\item Hyperparameter tuning: Hyperparameters play a crucial role in model performance. To find the optimal hyperparameters, one needs to try various hyperparameter combinations. DRAS requires no source code modifications to find hyperparameters. The customized hyperparameters can be passed to source code through command arguments. The configurable hyperparameters are ranging from learning rate, mini-batch size, epsilon, epsilon decay rate, and discount factor. This allows the developers to launch multiple simulations with different hyperparameters in parallel.
\item Performance comparison: The output of the simulation is in \verb|Results| folder by default and it includes the job and system information to evaluate scheduling performance. All simulations have the same output format regardless of scheduling policies. This enables a fair performance comparison of the different scheduling policies on the same log.
\item Extensibility: Modules are independent of each other. The scheduling policy plug-in only communicates with job scheduling module and is entirely decoupled from the rest of the system. DRAS employed the two most popular reinforcement learning algorithms, i.e., policy gradient and deep q-learning. In addition, there are several traditional scheduling algorithms, such as FCFS, SJF, and LJF (longest job first), to choose from. New scheduling algorithms can be implemented by creating a new scheduling policy plug-in and replacing the current plug-in. 
\item Rich debugging facility: We provides five-level logging options to record various events to \verb|Debug| folder. The most detailed logging information captures sufficiently detailed information allowing the developers to quickly identify the issues in their code. To achieve the best performance, the developers can set \verb|debug_lvl| argument to 1, which will record the minimum amount of information to reduce the I/O demand.
\end{itemize}

\section{Impact}\label{Impact}
Reinforcement learning is a highly active research field. Advanced reinforcement learning algorithms have been proposed in recent years \cite{mnih2016asynchronous,Silver,lillicrap2019continuous,schulman2017proximal,schulman2017trust}. However, not all reinforcement learning algorithms are suitable for our HPC scheduling problem. A good solution is supposed to achieve good scheduling performance, such as low average job wait time and high system utilization, with the minimum scheduling overhead. Typical HPC systems tolerate 10-30 seconds of scheduling delay \cite{Fan2, Topper}. It is crucial to evaluate the performance and overhead of new RL scheduling methods before deployment. Fortunately, our DRAS design can easily embrace new RL algorithms.

The underlying CQSim scheduling simulator has been successfully supporting a number of projects in this field over a decade \cite{Li1,Fan4,Fan3,Fan1,Fan5,xu01,Zheng1,Yang1,Fan6,Fan7,Fan8,Qiao1,Qiao2}. CQSim provides a unified platform to evaluate the performance of various methods with minimal overheads. For example, \cite{Fan5, Zheng1} used CQSim to explore advanced methods, such as reinforcement learning and plan-based methods, to scheduling HPC jobs. The advanced scheduling methods enable the HPC systems to achieve better user-level and system-level performance. \cite{Li1, Fan1, Fan8} utilized CQSim to explore various factors that could affect the performance of HPC job scheduling, such as job runtime estimate and system utilization. By identifying these factors, the system administrators could develop new policies to minimize the impacts of these factors. \cite{Fan4, Fan3, xu01,Fan6} aim to find the scheduling strategies to handle multiple resources, i.e., CPU, burst buffer, GPU, and power. CQSim plays a crucial role in these multi-resource projects, because CQSim simulator provides a virtual configurable platform to identify the best scheduling policy to schedule specific resources on a given system before deployment on real systems. \cite{Yang1, Qiao1,Qiao2} proposed and analyzed novel job placement algorithms on HPC system to improve the efficiency of job placement. Job placement problems are difficult to be measured and conducted on real systems due to the scale of the problems, CQSim provides an easy platform to evaluate the performance of various job placement solutions. Additionally, CQSim is a mature community with many existing scheduling policies, ranging from traditional utility-based policies, optimization methods, to popular RL methods. This enables quick and accurate performance comparison between the new and existing scheduling methods.

DRAS can not only serve the research purpose, more importantly, the trained RL model can directly deploy to real HPC systems. Thanks to the decoupled scheduling policy design, the system administrators could first train and evaluate the RL agent in the simulator using the historical job logs. Once they obtain satisfactory performance, they can transfer the trained RL agent to real HPC systems and schedule jobs in real-time.

\section{Conclusion and Future Work}\label{Conclusion and Future Work}
In this work, we present DRAS-CQSim, a reinforcement learning based HPC scheduling framework, which aims to train RL-based scheduling agents to outperform the traditional scheduling policies. In summary, DRAS-CQSim implements two RL-based scheduling methods, provides a standard platform to design and evaluate RL-based scheduling methods, and enables fair comparison of different scheduling methods. As further developments, we plan to implement and evaluate other RL methods, such as A2C, SAC, and PPO, in order to find the best RL-empowered method for HPC job scheduling.

\section*{Acknowledgements}
\label{}

This work is supported in part by US National Science Foundation grants CNS-1717763, CCF-1618776, and the U.S. Department of Energy, Office of Science, under contract DE-AC02-06CH11357, DE-AC02-05CH11231.

\section*{References}



\bibliographystyle{unsrt}
\bibliography{software.bib}





\section*{Required Metadata}
\label{}

\section*{Current code version}
\label{}
\begin{table}[H]
\begin{tabular}{|l|p{6.5cm}|p{6.5cm}|}
\hline
\textbf{Nr.} & \textbf{Code metadata description} & \textbf{Please fill in this column} \\
\hline
C1 & Current code version & v1.0  \\ 
\hline
C2 & Permanent link to code/repository used for this code version & $https://github.com/SPEAR-IIT/CQSim/tree/DRAS$ \\
\hline
C3  & Permanent link to Reproducible Capsule & https://codeocean.com/capsule/7855334/ \\
\hline
C4 & Legal Code License   & MIT \\
\hline
C5 & Code versioning system used & Git \\
\hline
C6 & Software code languages, tools, and services used & Python, Tensorflow, Keras \\
\hline
C7 & Compilation requirements, operating environments \& dependencies &  Any OS supporting Python \\
\hline
C8 & If available Link to developer documentation/manual & http://bluesky.cs.iit.edu/cqsim/ documents/Manual.pdf \\
\hline
C9 & Support email for questions & yfan22@hawk.iit.edu \\
\hline
\end{tabular}
\caption{Code metadata (mandatory)}
\label{} 
\end{table}

\section*{Current executable software version}
\label{}
\begin{table}[H]
\begin{tabular}{|l|p{6.5cm}|p{6.5cm}|}
\hline
\textbf{Nr.} & \textbf{(Executable) software metadata description} & \textbf{Please fill in this column} \\
\hline
S1 & Current software version &  v1.0 \\
\hline
S2 & Permanent link to executables of this version  & $https://github.com/SPEAR-IIT/CQSim/tree/DRAS$ \\
\hline
S3  & Permanent link to Reproducible Capsule & https://codeocean.com/capsule/7855334/ \\
\hline
S4 & Legal Software License & MIT \\
\hline
S5 & Computing platforms/Operating Systems & Any OS supporting Python \\
\hline
S6 & Installation requirements \& dependencies & Python, Tensorflow, Keras \\
\hline
S7 & If available, link to user manual - if formally published include a reference to the publication in the reference list & http://bluesky.cs.iit.edu/cqsim/ documents/Manual.pdf \\
\hline
S8 & Support email for questions &  yfan22@hawk.iit.edu \\
\hline
\end{tabular}
\caption{Software metadata (optional)}
\label{} 
\end{table}

\end{document}